\documentclass[preprint,12pt]{elsarticle}
\usepackage{graphics}

\usepackage{amssymb}

\journal{}

\begin{document}
\newcommand{\bqa}{\begin{eqnarray}}
\newcommand{\eqa}{\end{eqnarray}}
\newcommand{\nl}{\nonumber \\}
\def\db#1{\bar D_{#1}}
\def\d#1{D_{#1}}
\def\tld#1{\tilde {#1}}
\begin{frontmatter}



\title{Testing and improving the numerical accuracy of the NLO predictions}


\author{R. Pittau}

\address{Departamento de F\'isica Te\'orica y del Cosmos and CAFPE

Universidad de Granada, E-18971, Granada, Spain}

\begin{abstract}
I present a new and reliable method to test the numerical accuracy 
of NLO calculations
based on modern OPP/Generalized Unitarity techniques. 
A convenient solution to {\em rescue} most of the detected numerically 
inaccurate points is also proposed.
\end{abstract}


\end{frontmatter}


\section{Introduction}
With the advent of the modern OPP~\cite{Ossola:2006us,Ossola:2008xq} 
and Generalized Unitarity~\cite{Bern:1994zx,Bern:1994cg,Bern:1995db,Bern:1997sc,Britto:2004nc,Forde:2007mi,Ellis:2007br,Ellis:2008ir}
based techniques, the {\em art} of computing
NLO corrections received a lot of attention in the last few years and several 
programs~\cite{Ossola:2007ax,vanHameren:2009dr,Giele:2008bc,Berger:2008sj,Lazopoulos:2008ex,Mastrolia:2010nb} and 
computations~\cite{Bevilacqua:2009zn,Bevilacqua:2010ve,Ellis:2008qc,Ellis:2009zw,KeithEllis:2009bu,Berger:2009zg,Berger:2009ep} 
exist, by now, based on this philosophy.
 
 While for the traditional reduction methods~\cite{Passarino:1978jh} 
a lot of work has been 
spent already to find ways to control the numerical accuracy of the results 
~\cite{Denner:2005nn,Ellis:2005zh}, in the case of these new techniques the situation is still at an early stage.
However, their potential to {\em self detect} stability problems
is known since 2007~\cite{Ossola:2007ax}, the basic observation being that, since a reconstruction of a function $N(q)$ of the {\em would be} 
integration momentum $q$ is involved (the coefficients of which are 
interpreted as the coefficients of the scalar 1-loop functions entering the
calculation), one can numerically test 
the accuracy of it by comparing $N(q)$ and its re-constructed counterpart
at a new, arbitrarily chosen value of $q$. 
However, the arbitrariness of the point chosen for the test poses
serious problems, because it introduces a new, unwanted, parameter 
upon which the check depends in 
an unpredictable way~\footnote{Improvements on this technique have
 been recently presented in~\cite{Mastrolia:2010nb}.}.
Furthermore, not all the reconstructed coefficients enter into the actual
computation, because, for example, some of them may multiply vanishing 
loop functions, rendering immaterial a possible inaccuracy in their 
determination. In addition, it is not clear how to test the rational part 
of the amplitude~\cite{Ossola:2008xq}.

Nevertheless, it keeps being very tempting the idea of {\em self detecting} 
numerical inaccuracies, avoiding the need of additional analytic work. 
In this paper, I present a method to achieve this task, based 
on the construction of a reliable precision estimator working 
at an event by event basis.
Therefore, it becomes possible for the user to safely 
set a precision threshold above which the inaccurate points are discarded.
Furthermore, I prove that, re-fitting the discarded points
at higher precision {\em while keeping the computation
of $N(q)$ in double precision} allows to re-include most of them 
in the original sample. This solution nicely factorizes 
the problem, in the sense that the codes (of the parts of the code) 
computing the function $N(q)$ can be kept in double precision and  
only the fitting procedure to get the coefficients needs to be re-done
at higher precision.

The structure of this work is very simple:
in Sections~\ref{sec:meth} and~\ref{sec:meth1}, I describe the algorithm and, 
in Section~\ref{sec:tests}, I report on the tests I performed 
on the whole procedure.

\section{The method}
\label{sec:meth}
In the OPP technique the numerator $N(q)$ of
the integrand of a $m$-point amplitude is decomposed in terms of denominators 
$\d{i} = (q + p_i)^2-m_i^2$ \footnote{In our notation, $q$ in 4-dimensional, 
$\bar q$ $n$-dimensional and $n$-dimensional denominators are written as
$\db{i} = (\bar q + p_i)^2-m_i^2$.}
\bqa
\label{eq:opp}
N(q) &=& {\cal D}^{(m)}(q) +
\sum_{i_0 < i_1 < i_2 }^{m-1}
          c( i_0 i_1 i_2;q) 
\prod_{i \ne i_0, i_1, i_2}^{m-1} \d{i} 
      + 
\sum_{i_0 < i_1 }^{m-1}
          b(i_0 i_1; q) 
\prod_{i \ne i_0, i_1}^{m-1} \d{i} \nl
      & &+
\sum_{i_0}^{m-1}
          a(i_0; q) 
\prod_{i \ne i_0}^{m-1} \d{i}\,,
\eqa
where, for later convenience, I have grouped all the 4-point contributions into a single term
\bqa
\label{eq:oppd}
{\cal D}^{(m)}(q) &=& \sum_{i_0 < i_1 < i_2 < i_3}^{m-1}
          d( i_0 i_1 i_2 i_3; q) 
\prod_{i \ne i_0, i_1, i_2, i_3}^{m-1} \d{i} \,.
\eqa
The functions
$d( i_0 i_1 i_2 i_3; q)$, $c( i_0 i_1 i_2; q)$, $b( i_0 i_1; q)$ 
and $a(i_0; q)$ depend on the integration momentum $q$ and bring 
information on the coefficients of the scalar 1-loop integrals, that
are obtained  by fitting $N(q)$ at different 
values of $q$ that nullify, in turn, the denominators.
Finally, performing a global shift of all the masses appearing 
in the denominators of Eq.~\ref{eq:opp}
\bqa
\label{eq:shift}
m_i^2 \to m_i^2-\tld{q}^2
\eqa
and fitting again, allows to reconstruct also a piece 
of the rational terms, called $R_1$~\cite{Ossola:2008xq}.
In summary, by knowing the set \footnote{I use a notation such that
the coefficients of the scalar 1-loop functions have the same name of the functions appearing
in Eqs.~\ref{eq:opp} and ~\ref{eq:oppd}, but without $q$ dependence.}
\bqa
\label{eq:coeff1}
\begin{tabular}{lll}
$d( i_0 i_1 i_2 i_3)$\,, & $c( i_0 i_1 i_2)$\,, & \\
$b( i_0 i_1)$\,,         &  $a(i_0)$       \,, & $R_1$\,, 
\end{tabular}
\eqa
the amplitude $A$ is reconstructed by simply
multiplying by the corresponding
scalar 1-loop integrals~\cite{vanHameren:2009dr,Ellis:2007qk}
\footnote{The remaining piece of the 
rational terms $R_2$ 
can be computed  as explained 
in~\cite{Ossola:2008xq,Draggiotis:2009yb,Garzelli:2009is}.}.
\bqa
\label{eq:amp}
{A}  &=& 
\sum_{i_0 < i_1 < i_2 < i_3}^{m-1} d( i_0 i_1 i_2 i_3) 
\int d^n \bar{q} \frac{1}{\db{i_0}\db{i_1}\db{i_2}\db{i_3}} \nl
&&+ \sum_{i_0 < i_1 < i_2 }^{m-1}c( i_0 i_1 i_2)
\int d^n \bar{q} \frac{1}{\db{i_0}\db{i_1}\db{i_2}} \nl
&&+\sum_{i_0 < i_1 }^{m-1} b(i_0 i_1) 
\int d^n \bar{q} \frac{1}{\db{i_0}\db{i_1}} \nl
&&+ \sum_{i_0}^{m-1} a(i_0) \int d^n \bar{q} \frac{1}{\db{i_0}} +R_1\,. 
\eqa

The key point of the method I propose in this paper 
is the observation that, if one would be
able to obtain the whole set of Eq.~\ref{eq:coeff1}
in an independent way, giving, as a result, a new set 
\bqa
\label{eq:coeff2}
\begin{tabular}{lll}
$d^\prime( i_0 i_1 i_2 i_3)$\,, & $c^\prime( i_0 i_1 i_2)$\,, & \\
$b^\prime( i_0 i_1)$\,,         &  $a^\prime(i_0)$\,,        & $R_1^\prime$\,,
\end{tabular}
\eqa
an independent determination of the  1-loop amplitude would become possible
\bqa
\label{eq:amp1}
{A}^\prime &=& 
\sum_{i_0 < i_1 < i_2 < i_3}^{m-1} d^\prime( i_0 i_1 i_2 i_3) 
\int d^n \bar{q} \frac{1}{\db{i_0}\db{i_1}\db{i_2}\db{i_3}} \nl
&&+ \sum_{i_0 < i_1 < i_2 }^{m-1}c^\prime( i_0 i_1 i_2)
\int d^n \bar{q} \frac{1}{\db{i_0}\db{i_1}\db{i_2}} \nl
&&+\sum_{i_0 < i_1 }^{m-1} b^\prime(i_0 i_1) 
\int d^n \bar{q} \frac{1}{\db{i_0}\db{i_1}} \nl
&&+ \sum_{i_0}^{m-1} a^\prime(i_0) \int d^n \bar{q} \frac{1}{\db{i_0}}
 +R_1^\prime\,\,,
\eqa
that could then be used to define a reliable estimator of the accuracy
as follows \footnote{In an actual, numerical implementation, a 
small quantity $\epsilon$ has to be included 
in the denominator of Eq.~\ref{eq:esta} to deal with 
the case of vanishing amplitude.}
\bqa
\label{eq:esta}
E^A \equiv \frac{|A-A^\prime|}{|A|}\,. 
\eqa
The advantage of Eq.~\ref{eq:esta}, with respect to a test performed
at the level of the function $N(q)$, is that {\em only} the coefficients 
contributing to the amplitude enter into the game. 
Furthermore, a test on $R_1$ becomes possible.
As it will become clear shortly, it is convenient 
to differentiate the two cases where the coefficients 
of the sets in Eqs.~\ref{eq:coeff1} and ~\ref{eq:coeff2}  
(and, {\em a fortiori}, the amplitudes
in Eqs.~\ref{eq:amp} and ~\ref{eq:amp1}) are computed in double or 
multi-precision. Then, I denote the double precision estimator by
\bqa
\label{eq:estad}
E^A_d \equiv \frac{|A_d-A_d^\prime|}{|A_d|}\,,
\eqa
and its multi-precision version by
\bqa
\label{eq:estam}
E^A_m \equiv \frac{|A_m-A_m^\prime|}{|A_m|}\,. 
\eqa
In the rest of this section, I illustrate how to obtain the new set of 
Eq.~\ref{eq:coeff2}.

The technique is similar to the procedure adopted to compute $R_1$.
Under the shift in Eq.~\ref{eq:shift}, Eq.~\ref{eq:opp} 
becomes
\bqa
\label{eq:opp1}
N(q) &=& \bar {\cal  D}^{(m)}(q) +
\sum_{i_0 < i_1 < i_2 }^{m-1}
          \bar c( i_0 i_1 i_2;q) 
\prod_{i \ne i_0, i_1, i_2}^{m-1} (\d{i}+\tld{q}^2) \nl
      &&+ 
\sum_{i_0 < i_1 }^{m-1}
          \bar b(i_0 i_1; q) 
\prod_{i \ne i_0, i_1}^{m-1} (\d{i}+\tld{q}^2) \nl
      &&+
\sum_{i_0}^{m-1}
          \bar a(i_0; q) 
\prod_{i \ne i_0}^{m-1} (\d{i}+\tld{q}^2)\,,
\eqa
where
\bqa
\label{eq:expD}
\bar {\cal D}^{(m)}(q) &=& \sum_{j= 2}^{m} \tld{q}^{\,(2j-4)} d^{\,(2j-4)}(q)\,,
\eqa
with the last coefficient of $\bar {\cal D}^{(m)}(q)$  independent on $q$
\bqa
\label{eq:noq}
d^{\,(2m-4)}(q) = d^{\,(2m-4)}\,,
\eqa
and  where
\bqa
\label{eq:qt2dep}
\bar c( i_0 i_1 i_2;q) &=& c( i_0 i_1 i_2;q) 
  + \tld{q}^2 c^{\,(2)}( i_0 i_1 i_2;q) \nl
\bar b( i_0 i_1;q) &=& b( i_0 i_1;q) + \tld{q}^2 b^{\,(2)}( i_0 i_1;q) \nl
\bar a( i_0;q) &=& a( i_0;q)\,.
\eqa
This last equation implies, for the 1-,2 and 3-point coefficients
\bqa
\label{eq:sys1}
\bar c( i_0 i_1 i_2) &=& c( i_0 i_1 i_2) 
  + \tld{q}^2 c^{\,(2)}( i_0 i_1 i_2) \nl
\bar b( i_0 i_1) &=& b( i_0 i_1) + \tld{q}^2 b^{\,(2)}( i_0 i_1) \nl 
\bar a( i_0) &=& a( i_0)\,.
\eqa
The constants $b^{\,(2)}( i_0 i_1)$, $c^{\,(2)}( i_0 i_1 i_2)$ and 
$d^{(2m-4)}$ enter into the computation of $R_1$~\cite{Ossola:2008xq}
\bqa
\label{erre1}
{R}_1 &=& - \frac{i}{96 \pi^2} d^{(2m-4)}
      -  \frac{i}{32 \pi^2}
\sum_{i_0 < i_1 < i_2 }^{m-1}
          c^{(2)}( i_0 i_1 i_2)
 \nl
     &-& \frac{i}{32 \pi^2}
\sum_{i_0 < i_1 }^{m-1}
          b^{(2)}(i_0 i_1)
\left(m_{i_0}^2+m_{i_1}^2- \frac{(p_{i_0}-p_{i_1})^2}{3}\right )\,, 
\eqa
and can be determined with the help of Eqs.~\ref{eq:sys1} and ~\ref{eq:expD}. 

With the knowledge of  
$a(i_0)$, $b(i_0 i_1)$, $c( i_0 i_1 i_2)$, $b^{(2)}(i_0 i_1)$ and
$c^{(2)}( i_0 i_1 i_2)$, $a^\prime$, $b^\prime$ 
and $c^\prime$  in Eq.~\ref{eq:coeff2} can be immediately obtained
with a new mass shift
\bqa
\label{eq:shift1}
m_i^2 \to m_i^2-\tld{q}_1^{2}\,,
\eqa
giving
\bqa
\label{eq:sys2}
\bar c_1( i_0 i_1 i_2) &=& c( i_0 i_1 i_2) 
  + \tld{q}_1^2 c^{\,(2)}( i_0 i_1 i_2) \nl
\bar b_1( i_0 i_1) &=& b( i_0 i_1) + \tld{q}_1^2 b^{\,(2)}( i_0 i_1) \nl 
\bar a_1( i_0) &=& a( i_0)\,,
\eqa
where I attached the subscript $_1$ to the coefficients obtained
with the new shift. Combining Eqs.~\ref{eq:sys1} and~\ref{eq:sys2} gives
\bqa
\label{eq:res1}
 a^\prime( i_0) &=& \bar a_1( i_0)\, \nl
 b^\prime( i_0 i_1) &=& 
 \frac{\bar b( i_0 i_1) + \bar b_1( i_0 i_1)}{2}
-\frac{\tld{q}^{2} + \tld{q}_1^{2}}{2}\,\, b^{\,(2)}( i_0 i_1)\,, \nl
 c^\prime( i_0 i_1 i_2) &=& 
 \frac{\bar c( i_0 i_1 i_2) + \bar c_1( i_0 i_1 i_2)}{2}
-\frac{\tld{q}^{2} + \tld{q}_1^{2}}{2}\,\, c^{\,(2)}( i_0 i_1 i_2)\,.
\eqa
As for $R_1$, an independent determination of
$b^{\,(2)}( i_0 i_1)$, $c^{\,(2)}( i_0 i_1 i_2)$ in Eq.~\ref{erre1} 
also follows from the new shift
\bqa
\label{eq:res2}
 c^{\prime\,(2)}( i_0 i_1 i_2) &=& 
 \frac{\bar c( i_0 i_1 i_2) -
       \bar c_1( i_0 i_1 i_2)}{\tld{q}^{2}-\tld{q}_1^{2}} \nl
 b^{\prime\,(2)}( i_0 i_1 ) &=& 
 \frac{\bar b( i_0 i_1 ) -
       \bar b_1( i_0 i_1 )}{\tld{q}^{2}-\tld{q}_1^{2}}\,. 
\eqa
With the help of Eq.~\ref{eq:sys2} one can now
completely reconstruct the 1-, 2- and 3-point parts of the numerator 
function with masses shifted according to Eq.~\ref{eq:shift1}, namely
\bqa
\label{eq:piece}
&&\sum_{i_0 < i_1 < i_2 }^{m-1}
          \bar c_1( i_0 i_1 i_2;q) 
\prod_{i \ne i_0, i_1, i_2}^{m-1}(\d{i}+\tld{q}_1^{2}) 
      + 
\sum_{i_0 < i_1 }^{m-1}
          \bar b_1(i_0 i_1; q) 
\prod_{i \ne i_0, i_1}^{m-1} (\d{i}+ \tld{q}_1^{2}) \nl
      &&+
\sum_{i_0}^{m-1}
         \bar a_1(i_0; q) 
\prod_{i \ne i_0}^{m-1} (\d{i}+\tld{q}_1^{2}) \,.
\eqa
By subtracting Eq.~\ref{eq:piece} from $N(q)$ one determines 
$\bar {\cal D}_1^{(m)}(q)$ obeying the following polynomial 
(in  $\tld{q}_1^2$) representation (see Eq.~\ref{eq:expD})
\bqa
\label{eq:newD}
\bar {\cal D}_1^{(m)}(q) &=& 
\sum_{j= 2}^{m} \tld{q}_1^{\,(2j-4)} d_1^{\,(2j-4)}(q)\,,
\eqa
the first coefficient of which, computed at values
$q= q_{i_0,i_1,i_2,i_3}$ nullifying, in turns, all possible combinations
of 4 denominators
\bqa
\d{i_0}(q_{i_0,i_1,i_2,i_3})
=\d{i_1}(q_{i_0,i_1,i_2,i_3}) 
=\d{i_2}(q_{i_0,i_1,i_2,i_3}) =\d{i_3}(q_{i_0,i_1,i_2,i_3}) = 0\,,
\eqa
gives the desired independent determination of the box 
coefficients
\bqa
\label{eq:newbox}
d^\prime( i_0 i_1 i_2 i_3)= d_1^{\,(0)}(q_{i_0,i_1,i_2,i_3})\,.
\eqa
From the last term in Eq.~\ref{eq:newD} one obtains, instead 
\bqa
\label{eq:newd}
d^{\prime\,(2m-4)}= d_1^{\,(2m-4)}\,,
\eqa
that, together with the coefficients
in Eq.~\ref{eq:res2}, gives a complete alternative 
determination of $R_1$
\bqa
\label{erre1p}
{R}_1^\prime &=& - \frac{i}{96 \pi^2} d^{\prime\,(2m-4)}
      -  \frac{i}{32 \pi^2}
\sum_{i_0 < i_1 < i_2 }^{m-1}
          c^{\prime\,(2)}( i_0 i_1 i_2)
 \nl
     &-& \frac{i}{32 \pi^2}
\sum_{i_0 < i_1 }^{m-1}
          b^{\prime\,(2)}(i_0 i_1)
\left(m_{i_0}^2+m_{i_1}^2- \frac{(p_{i_0}-p_{i_1})^2}{3}\right )\,, 
\eqa

I close the Section by summarizing the procedure. 
One fits the numerator function $N(q)$ three times; the first time 
with $\tld{q}^2= 0$ (Eq.~\ref{eq:opp}) to determine the
cut-constructible part of the amplitude, namely all of the coefficients 
in Eq.~\ref{eq:coeff1}; the second time with  $\tld{q}^2 \ne 0$, to compute
$R_1$ by means of Eq.~\ref{erre1}; the third time
with a new value of the mass shift ($\tld{q}_1^{\,2}$) to calculate
the alternative set of coefficients 
in Eqs.~\ref{eq:res1},~\ref{eq:res2},~\ref{eq:newbox} and
~\ref{eq:newd}, that allow to
build the precision estimator $E^A_d$   given in Eq.~\ref{eq:estad}.

It is clear that performing
the three fits by using directly the numerator 
function $N(q)$ appearing in the l.h.s. of Eq.~\ref{eq:opp}
could be computationally very expensive because,
in practical cases, the calculation of $N(q)$ is rather time consuming
\footnote{I assume here an 
unoptimized computation of $N(q)$, performed without cashing the information
that does not depend on $q$.}. Fortunately, after the first fit, 
one is allowed to  use the {\em reconstructed}
numerator function (namely the r.h.s. of Eq.~\ref{eq:opp}) to determine
both $R_1$ and $E^A_d$. The additional CPU time is then very moderate. 
To further decrease it, one can also observe that, instead of 
determining $d^{\,(2m-4)}$ through the expansion in
Eq.~\ref{eq:expD}, that requires the knowledge of $\bar {\cal D}^{(m)}(q)$ 
at $(m-2)$ different values of $\tld{q}^2$, one can get it
by means of the following relation among the OPP coefficients
\bqa
\label{eq:bewd4}
d^{\,(2m-4)}
  =
 &-& 
\sum_{i_0 < i_1 < i_2 }^{m-1}
           c^{(2)}( i_0 i_1 i_2;q) 
 -
\sum_{i_0 < i_1 }^{m-1}
           b(i_0 i_1; q) 
 -
\sum_{i_0 < i_1 }^{m-1}
           b^{(2)}(i_0 i_1; q) 
\sum_{i \ne i_0, i_1}^{m-1} \d{i} \nl
 &-&
\sum_{i_0}^{m-1}
          a(i_0; q) 
\sum_{i \ne i_0}^{m-1} \d{i} \,.
\eqa
Eq.~\ref{eq:bewd4} is proved in~\ref{appa}.

\section{Rescuing the inaccurate points}
\label{sec:meth1}
Under the assumption that $E^A_d$ in Eq.~\ref{eq:estad} is a good 
precision estimator, 
points can be rejected when $E^A_d > E_{\lim}$, where
$E_{\lim}$ is a threshold value chosen by the user.
 In this Section, I propose, as a simple recipe to {\em rescue} the
rejected points,
to re-perform the three fits described in Section \ref{sec:meth}
at higher precision {\em while keeping the computation
of $N(q)$ in double precision}. The advantage of this recipe  
is that multi-precision routines need to be implemented {\em just in the fitting
program}, while the code providing $N(q)$ 
can be left untouched \footnote{This strategy is especially relevant in 
the case of programs that already implement, internally, 
multi-precision routines~\cite{multilib}, such as {\tt CutTools}~\cite{Ossola:2007ax}.}. 
A new test in multi-precision can then be performed 
on the rescued points using the multi-precision estimator of Eq.~\ref{eq:estam}
and only if, even in this case, $E^A_m > E_{\lim}$ the event is discarded 
for good (or re-computed, if possible, with $N(q)$ also evaluated in 
multi-precision).
The hope is that the percentage of points rejected by this second 
test is very limited, so that they can be safely eliminated from the sample.
The effectiveness of this strategy is studied in the next Section.

\section{Testing the method}
\label{sec:tests}
To test the procedures described in the previous two Sections 
I implemented in {\tt CutTools}, as a numerator function $N(q)$
mimicking the full amplitude $A$, one of the 120 diagrams 
contributing to the $\gamma \gamma \to 4 \gamma$
scattering in massless QED \footnote{This diagram contains up to rank six
6-point functions, so it fairly represents the complexity 
of the real situations.}. To enhance the problematic region I did 
not apply any cut on the final state particles, so that numerically unstable
Phase Space configurations with zero Gram determinant can be freely approached.
From a practical point of view, I constructed the alternative amplitude
of Eq.~\ref{eq:amp1} by keeping the 4-point coefficients and re-computing only 
$c^\prime$, $b^\prime$, $a^\prime$ and $R_1^\prime$ with the help of
Eqs.~\ref{eq:res1},~\ref{eq:res2} and \ref{eq:newd}. The reason is that, 
in practice, the derivation of
the 4-point part of an amplitude is numerically quite stable, 
the bulk of the numerical instabilities coming from the lower-point sectors.

Before describing in details the tests, I introduce, 
besides the estimators given in  Eqs.~\ref{eq:estad} and Eqs.~\ref{eq:estam}, 
a few more variables.
I define two {\em true precision variables} as follow
\bqa
\label{eq:trueprec}
P_d=|A_d-A_e|/|A_e|~~~{\rm and}~~~P_m=|A_m-A_e|/|A_e|\,,
\eqa 
where $A_e$ is the {\em exact} reference amplitude computed 
with both fits {\em and} $N(q)$ in multi-precision.
$P_d$ and $P_m$ will be used, in the following, 
to test the actual precision in the computation of $A_d$ and $A_m$.
Furthermore, for the sake of comparison, I define two additional precision 
estimators, based on the so called  $N=N$ test of \cite{Ossola:2007ax}
\bqa
E^N_d= |N_d-N_{d,rec}|/|N_d|~~~{\rm and}~~~E^N_m= |N_d-N_{m,rec}|/|N_d|\,,
\eqa
where $N_d$ is the numerator function $N(q)$ computed, in double precision,
at a random value of $q$ \footnote{I picked up the point
$q= \sqrt{s}\,(1/2,-1/3,1/4,-1/5)$.},  
$N_{d,rec}$ the same numerator, but reconstructed, in double precision, via the r.h.s. of Eq.~\ref{eq:opp},
and, finally, $N_{m,rec}$ is the numerator function reconstructed by means of
a multi-precision fitting procedure.

\begin{figure}[t]
\begin{center}
\includegraphics[scale=0.5,angle=90]{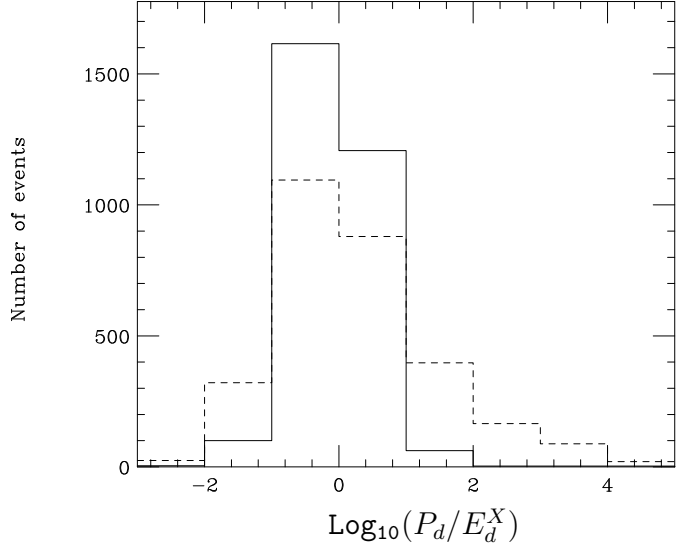} \\
\hspace{2.cm} ${\tt Log_{10}}(P_d/E^{X}_d)$
\caption{\label{fig_a}
Distribution of the ratio between the true precision 
$P_d= |A_d-A_e|/|A_e|$ and two different precision
estimators. The dashed histogram refers to the estimator at the numerator level
$E^{X}_d= E^{N}_d= |N_d-N_{d,rec}|/|N_d|$, 
the solid one to the estimator at the 
amplitude level $E^{X}_d= E^{A}_d=|A_{d}-A^\prime_{d}|/|A_{d}|$.}
\end{center}
\end{figure}

\begin{figure}[t]
\begin{center}
\includegraphics[scale=0.5,angle=90]{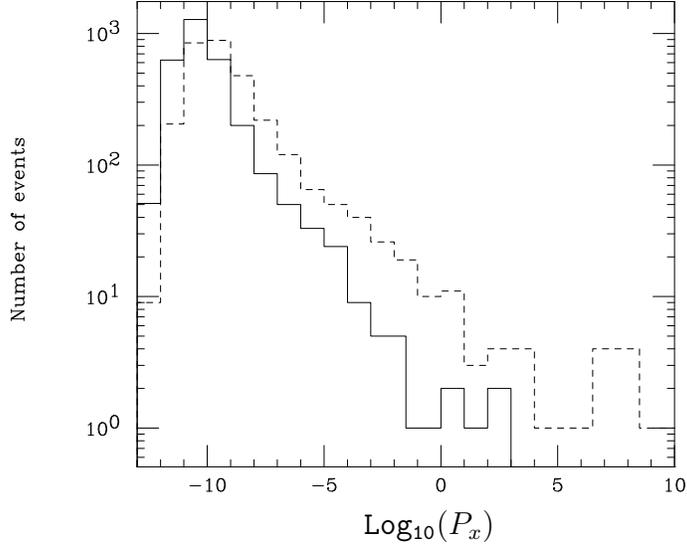}\\
\hspace{2cm} ${\tt Log_{10}}(P_x)$
\caption{\label{fig_b}
Distribution of the true precision variables 
$P_x=|A_x-A_e|/|A_e|$ (see text).
The dashed histograms refers to the double precision result
($P_x=P_d$), the solid histogram to the case with
fitting procedure carried out in multi-precision, but numerator function
computed in double precision ($P_x=P_m$).}
\end{center}
\end{figure}

\begin{figure}
\begin{center}
\includegraphics[scale=0.35,angle=90]{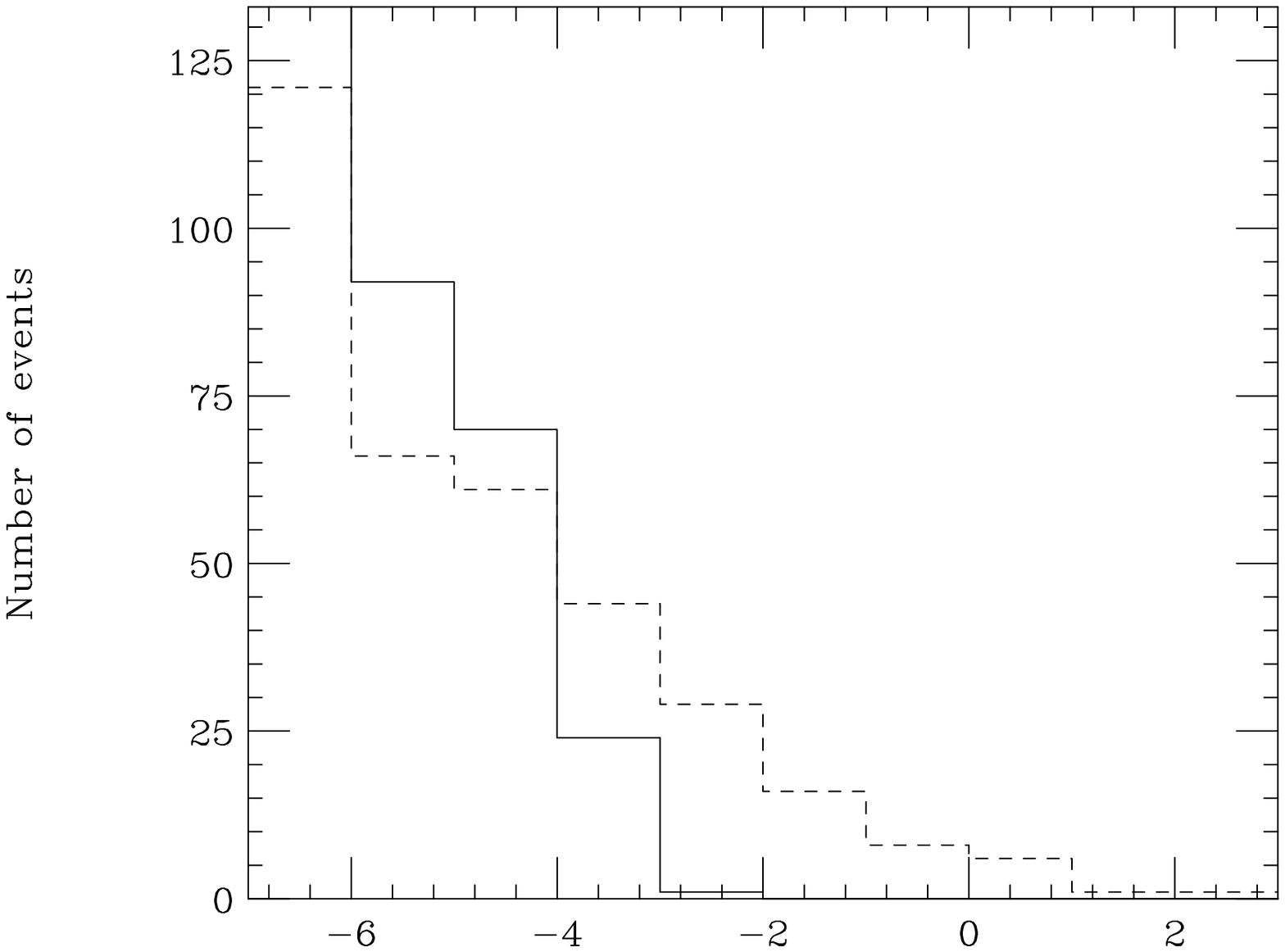} \hspace{0.2cm}
\includegraphics[scale=0.35,angle=90]{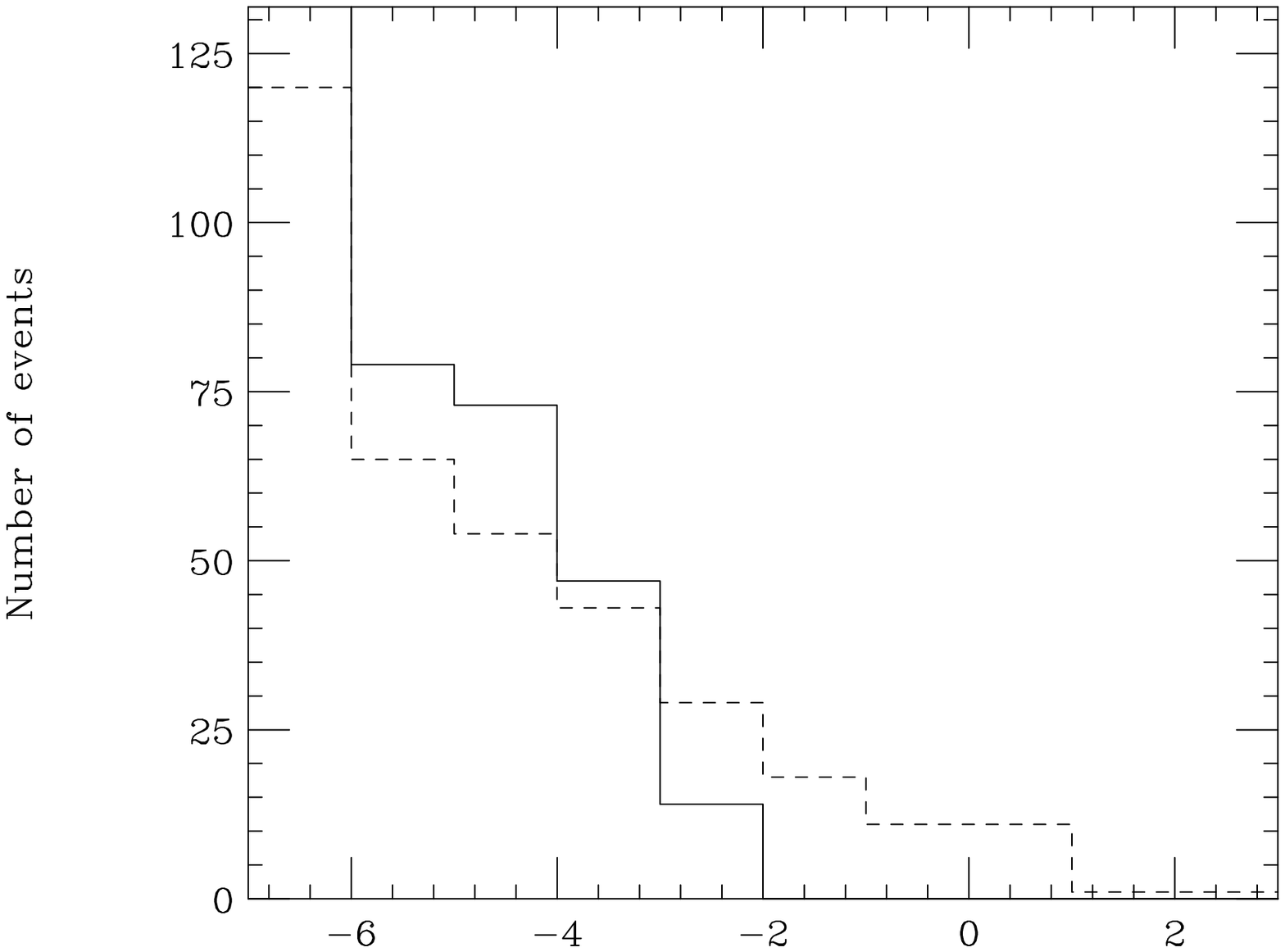}\\
\end{center}

\vspace{-5cm} 

\hspace{4cm} 
{\small $E_{\rm lim} = 10^{-4}$} \hspace{4cm}
{\small $E_{\rm lim} = 10^{-3}$}

\vspace{4cm} 
\hspace{2.5cm} 
\hspace{0.8cm} ${\tt Log_{10}}(P_d)$
\hspace{4.5cm} ${\tt Log_{10}}(P_d)$
\begin{center}
\includegraphics[scale=0.35,angle=90]{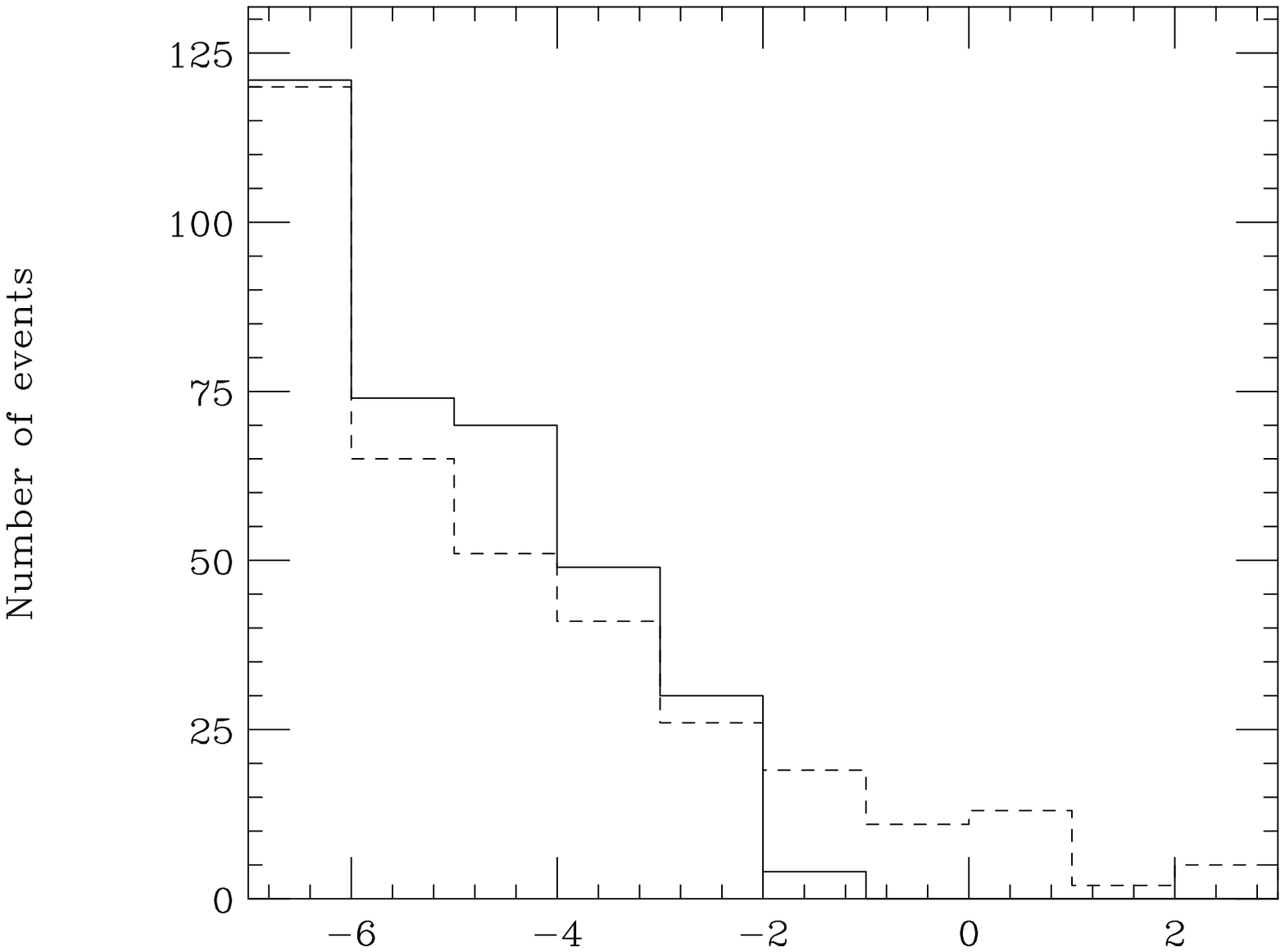} \hspace{0.2cm}
\includegraphics[scale=0.35,angle=90]{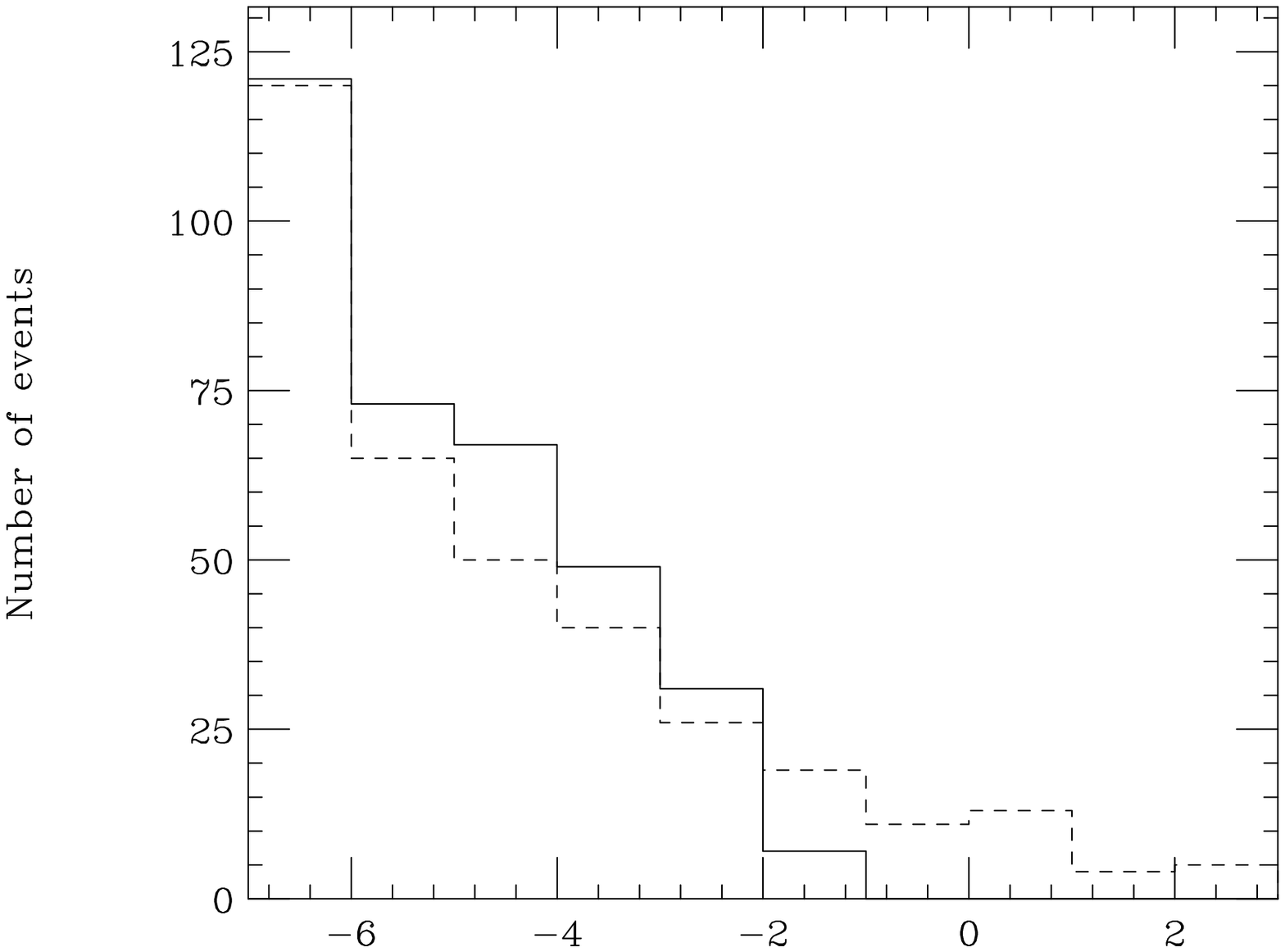}\\
\end{center}

\vspace{-4.8cm} 

\hspace{3.cm} 
{\small $E_{\rm lim} = 0.5 \times 10^{-2}$} \hspace{4cm}
{\small $E_{\rm lim} = 10^{-2}$}

\vspace{4cm} 
\hspace{2.5cm} 
\hspace{0.8cm} ${\tt Log_{10}}(P_d)$
\hspace{4.5cm} ${\tt Log_{10}}(P_d)$
\begin{center}
\caption{ \label{fig_14}
The tails of the distributions of the true precision variable 
$P_d$, with an additional constraint on the value 
of the precision estimators $E^{A}_d$, $E^{A}_m$, $E^{N}_d$ and 
$E^N_m= |N_d-N_{m,rec}|/|N_d|$ (see text). 
In the solid histograms, when $E^{A}_d > E_{\rm lim}$, a 
{\em rescue} of the point is performed by re-fitting the 1-loop coefficients
in multi-precision 
(with numerator functions kept in 
double precision) and, if also $E^{A}_m > E_{\rm lim}$, the event is discarded. 
In the dashed histograms, the same procedure is applied, but using the
estimators $E^{N}_d$ and $E^{N}_m$ instead.}
\end{center}
\end{figure}

In Figs.~\ref{fig_a}-\ref{fig_14}, 
I collect the results obtained by using 3000 random, uniformly distributed
Phase Space points.
In Fig.~\ref{fig_a} I plot the distributions of the ratios
$P_d/E^{N}_d$ and $P_d/E^{A}_d$. In the latter case (solid histogram), 
most of points fall within 2 orders of magnitude, indicating that 
$E^{A}_d$ is expected to accurately estimate the {\em true} numerical 
precision. This is not the case for the estimator based on the $N=N$ test.
The long right tail in the $P_d/E^{N}_d$ distribution (dashed histogram) shows 
that there are points for which the accuracy is badly overestimated
by $E^{N}_d$.
In  Fig.~\ref{fig_b}, I plot the distributions of $P_d$ (dashed histogram) 
and $P_m$ (solid histogram). It can be seen that keeping the computation 
of $N(q)$ in double precision, while performing the fit 
in multi-precision improves the accuracy. Nevertheless, points exist
for which $A_m$ is still not accurate enough, even if the 
solid plot stops order of magnitudes before the dashed one.  
In Fig.~\ref{fig_14}, I show the tails of the $P_d$ distribution, 
imposing four different cuts in the value of the estimators $E^A_d$ and
$E^A_m$. In the solid histograms, when $E^{A}_d > E_{\rm lim}$, a 
{\em rescue} of the point is performed by re-fitting the 1-loop coefficients
in multi-precision, while keeping the computation of the
numerator function in double precision, and, if also $E^{A}_m > E_{\rm lim}$, 
the event is discarded. 
In the dashed histograms, the same procedure is applied, but using, as
estimators, $E^{N}_d$ and $E^{N}_m$. Again, the right tails of the 
dashed histograms show that $E^{N}_d$ and $E^{N}_m$ are not good estimators
of the numerical accuracy, while the absence of points above $10^{-1}$ 
in the case of all the solid plots, indicates that $E^{A}_d$ and $E^{A}_m$ are 
able to select the {\em bad} inaccurate points quite efficiently.
For reader's reference I summarize, in table~\ref{tab:tab1}, a statistics 
of the number of points computed in multi-precision and 
discarded in each of the 4 cases.
As a conclusion, the {\rm rescue} procedure is able to recover 
most of them.
 
\begin{table}[t]
\begin{center}
\begin{tabular}{|c|l|l|}
\hline
$E_{lim}$  & $N_{mp}$ &  $N_{dis}$   \\
\hline 
\hline 
$10^{-4}$ & 90&  14 \\
\hline 
$10^{-3}$ & 62 & 8 \\
\hline 
$.5 \times 10^{-2}$ & 44& 6 \\
\hline 
$10^{-2}$ & 40 & 6 \\
\hline 
\end{tabular}
\end{center}
\caption{
\label{tab:tab1}
The number of points computed in multi-precision ($N_{mp}$) thanks 
to the {\em rescue} procedure
and the number of points discarded ($N_{dis}$) as a function of the threshold
value $E_{lim}$. The numbers refers to the solid histograms of 
Fig.~\ref{fig_14}, over a total number of 3000 events.}
\end{table}

As a final check on the goodness of the estimator 
at the amplitude level, I present, in table~\ref{tab:tab2}, 
the quantity $\max[ {\tt Log_{10}}(P_d)]-{\tt Log_{10}}(E_{lim})$ 
as a function of $E_{lim}$, when using the precision estimator $E^A_d$.
This variable measures the difference between the worst detected point,
in an analysis like that one represented by the solid histograms 
of Fig. \ref{fig_14}, and the chosen threshold value $E_{lim}$ for $E^A_d$. 
It can be seen that, for values of $E_{lim}$ 
between $10^{-2}$ and $10^{-6}$, $E^A_d$ overestimates the accuracy at most
by $1.1$ decimals and that the points where the overestimate is by almost
$2$ decimals lie in the safe region of very small values of $E_{lim}$, from
which one argues that $E^A_d$ is able to detect badly instable Phase 
Space points in a reliable way.

\begin{table}[t]
\begin{center}
\begin{tabular}{|c|c|}
\hline
$E_{lim}$  &  $\max[{\tt Log_{10}}(P_d)]-{\tt Log_{10}}(E_{lim})$   \\
\hline 
\hline 
$10^{-2}$ &              0.80         \\
\hline 
$5 \times 10^{-3}$ &     1.1         \\
\hline 
$10^{-3}$ &              0.74         \\
\hline 
$5 \times 10^{-4}$ &     0.57         \\
\hline 
$10^{-4}$ &              1.1         \\
\hline 
$5 \times 10^{-5}$ &     0.98         \\
\hline 
$10^{-5}$ &              1.0         \\
\hline 
$5 \times 10^{-6}$ &     1.1         \\
\hline 
$10^{-6}$ &              1.0        \\
\hline 
$10^{-8}$ &              1.7        \\
\hline 
$10^{-10}$&              1.8         \\
\hline 
\end{tabular}
\end{center}
\caption{The variable $\max[{\tt Log_{10}}(P_d)]-{\tt Log_{10}}(E_{lim})$ as a function
of $E_{lim}$, when using the precision estimator $E^A_d$.
\label{tab:tab2}}
\end{table}

\section{Conclusions}
I introduced a novel method to test the numerical accuracy 
of the NLO results produced by modern OPP/Generalized Unitarity techniques. 
The key ingredient is a re-computation of the 1-loop coefficients based 
on the properties of the OPP equation under a global shift of all the masses.
This re-computation can be performed by using the function previously reconstructed during 
the determination of the cut-constructible part of the amplitude, therefore 
at a moderate CPU time cost.
As a by-product, I also introduced a faster determination of one of the
coefficients contributing to the rational piece of the amplitude.
I proved, with numerical tests, the reliability of the procedure
and I proposed a convenient solution to {\em rescue} most of the 
detected numerically inaccurate points in a way 
that allows the computation of the integrand to remain in double precision.

\section{Acknowledgments}
Work supported by the European Community under contract MRTN-CT-2006-035505
and by the Spanish MEC under project FPA2008-02984.

\appendix
\section{An alternative determination of $d^{\,(2m-4)}$}
\label{appa}
The last coefficient $d^{\,(2m-4)}$ in the expansion of Eq.~\ref{eq:expD} contributes to $R_1$ through Eq.~\ref{erre1}.
In this appendix, I present a novel technique to determine it
from the other, known, coefficients of the OPP expansion.

The starting points are Eqs.~\ref{eq:opp1}-\ref{eq:qt2dep}. The l.h.s.
of Eq.~\ref{eq:opp1} does not depend on $\tld{q}^2$, so that one can equate 
to zero the coefficients of all the powers of $\tld{q}^2$ 
appearing in the r.h.s.
From the highest power, $\tld{q}^{\,(2m-2)}$, one obtains 
\bqa
\sum_{i_0 < i_1 }^{m-1}
          b^{(2)}(i_0 i_1; q) 
 +    
\sum_{i_0}^{m-1}
          a(i_0; q) = 0\,,
\eqa
while the next to highest power, $\tld{q}^{\,(2m-4)}$, gives~Eq.~\ref{eq:bewd4}.
Notice that the r.h.s. of Eq.~\ref{eq:bewd4} can be computed 
at arbitrary values of $q$, allowing extra numerical checks.

\end{document}